%% file: paper.tex
\documentclass[sigconf]{acmart}

\usepackage{lipsum}

\usepackage{xifthen}
\usepackage{multirow}
\usepackage{verbatim}
\usepackage{amsmath}



\usepackage{subcaption}
\PassOptionsToPackage{hyphens}{url}
\usepackage[hyphens]{url}
\usepackage{hyperref}
\usepackage{booktabs}
\usepackage{graphicx}
\usepackage{caption}
\usepackage{pdfpages}
\usepackage{setspace}
\usepackage{subfiles}
\usepackage[font=small,labelfont=bf]{caption}
\usepackage{url}

\date{} 

\setcopyright{none}
\renewcommand\footnotetextcopyrightpermission[1]{} 
\begin{document}






%


\title{Characterization and Comparison of Application Resilience for Serial and Parallel Executions}  
\vspace{-20pt}

\author{Kai Wu}
\email{kwu42@ucmerced.edu}
\affiliation{
  \institution{U. of California, Merced}
}

\author{Qiang Guan, Nathan DeBardeleben}
\email{{qguan, ndebard}@lanl.gov}
\affiliation{%
  \institution{USRC, Los Alamos National Lab}
}


\author{Dong Li}
\email{dli35@ucmerced.edu}
\affiliation{%
  \institution{U. of California, Merced}
}
\vspace{-20pt}

   \vspace{-20pt}
\begin{abstract}  
   \vspace{-5pt}

Soft error of exascale application is a challenge problem in modern High Performance Computing. In order to quantify an application’s resilience and vulnerability, the application-level fault injection method is widely adopted by HPC users. However, it is not easy since users need to inject a large number of faults to ensure statistical significance, especially for parallel version program. Normally, parallel execution is more complex and requires more hardware resources than its serial execution. Therefore, it is essential that we can predict error rate of parallel application based on its corresponding serial version. In this poster, we characterize fault pattern in serial and parallel executions. We find first there are same fault sources in serial and parallel execution. Second, parallel execution also has some \textit{\textbf{unique}} fault sources compared with serial executions. Those \textit{\textbf{unique}} fault sources are important for us to understand the difference of fault pattern between serial and parallel executions. 
   \vspace{-5pt}

\end{abstract}
   \vspace{-20pt}

\maketitle

\renewcommand{\thefootnote}{\fnsymbol{footnote}}
\footnotetext[1]{LA-UR-17-26470}

\input text/introduction
\input text/eval_methodology
\input text/evaluation_results

\input text/conclusions

%



%
\bibliographystyle{abbrvnat}
\bibliography{kai.bib}  

%
\end{document}

%% file: text/introduction.tex
   \vspace{-12pt}
\section{Introduction}
   \vspace{-5pt}
Making system resilient to hardware and software faults is a critical design goal for future extreme scale systems. 
To implement resilient HPC, we must have a good understanding of application resilience in the existence of faults. 
Currently, the application level fault injection is the major method to understand application resilience.
The application level fault injection triggers random bit flip in the operand or result of a random instruction. 
Typically, the statistical results of many fault injection tests, e.g., the percentage of the fault injection tests that have success application outcome, is used to evaluate the application resilience.

However, the application level fault injection can be very expensive, because HPC users need to inject a large number of faults to ensure statistical significance. Moreover, comparing with fault injection for the serial execution, fault injection for the parallel execution can be even more expensive. First, the parallel version needs more hardware resource than the serial version to deploy fault injection tests. Second, injecting faults into the parallel execution can be more difficult,  because there is a larger exploration space for fault injection.

In this poster, we explore the correlation between the parallel and serial executions regarding their resilience. Our ultimate goal is that by studying the resilience of the serial execution we can derive the resilience of the parallel execution without using expensive fault injection. We aim to answer two fundamental questions. First, does the application resilience remain the same across the serial and parallel executions? Second, if the application resilience is difference between the two executions,  what code structure causes such difference? We use an application-level fault injection tool named PFSEFI~\cite{PFSEFI:SIMUTOOLS16} to randomly choose dynamic instruction and then randomly flip one bit in the instruction result. After enough fault injection tests, we characterize and compare the serial and parallel execution codes based on the fault injection results. 
We hope that our work can lay foundation to build a model to predict the resilience of the parallel execution only based on fault injection results in the serial execution.
   \vspace{-10pt}

%% file: text/eval_methodology.tex
\vspace{-3pt}
\section{Evaluation Methodology}
   \vspace{-5pt}
We employ a fault injection tool, PFSEFI to study three NAS benchmarks (CG, FT, BT) with the input problem $S$. 
For serial execution fault injection, we only run one MPI process; For parallel execution fault injection, we run four MPI processes and then randomly choose one MPI process for fault injection. We inject faults into the whole application and focus on two types of instructions, i.e., floating point addition (\textit{fadd}) and floating point multiplication (\textit{fmul}), because they are the most common ones in HPC applications. To ensure statistical significance for fault injection, we gradually increase the number of fault injection tests until the fault injection result becomes stable. 
The fault injection results are classified into three types: (1) Benign: the computation results of benchmarks pass the benchmarks' verification phase, it means the computation results are acceptable. But the computation results may be different from those without fault injection. (2) Silent data corruption (SDC): the computation results of benchmarks do not pass the benchmarks' verification phase; (3) Crashes: the benchmark cannot run to completion. Since the fault injection happens based on the random selection of dynamic instruction, we cannot know where the fault happens within the application code. But we can know  the instruction address in the EIP register when the fault happens. 
We map the instruction address into the application code via PYELFTOOLS~\cite{pyelftool:github}.
Based on the EIP information for all random fault injection points, we can know the occurrence frequency of each faulty instruction; also, we can analyze the code, and understand the difference or similarity of application resilience in serial and parallel executions.

   \vspace{-10pt}

%% file: text/evaluation_results.tex

\vspace{-2pt}
   
\section{EXPERIMENT RESULTS}
   \vspace{-5pt}

\begin{figure}[!t]
    \centering
    \includegraphics[width=0.46\textwidth, height=0.15\textheight]{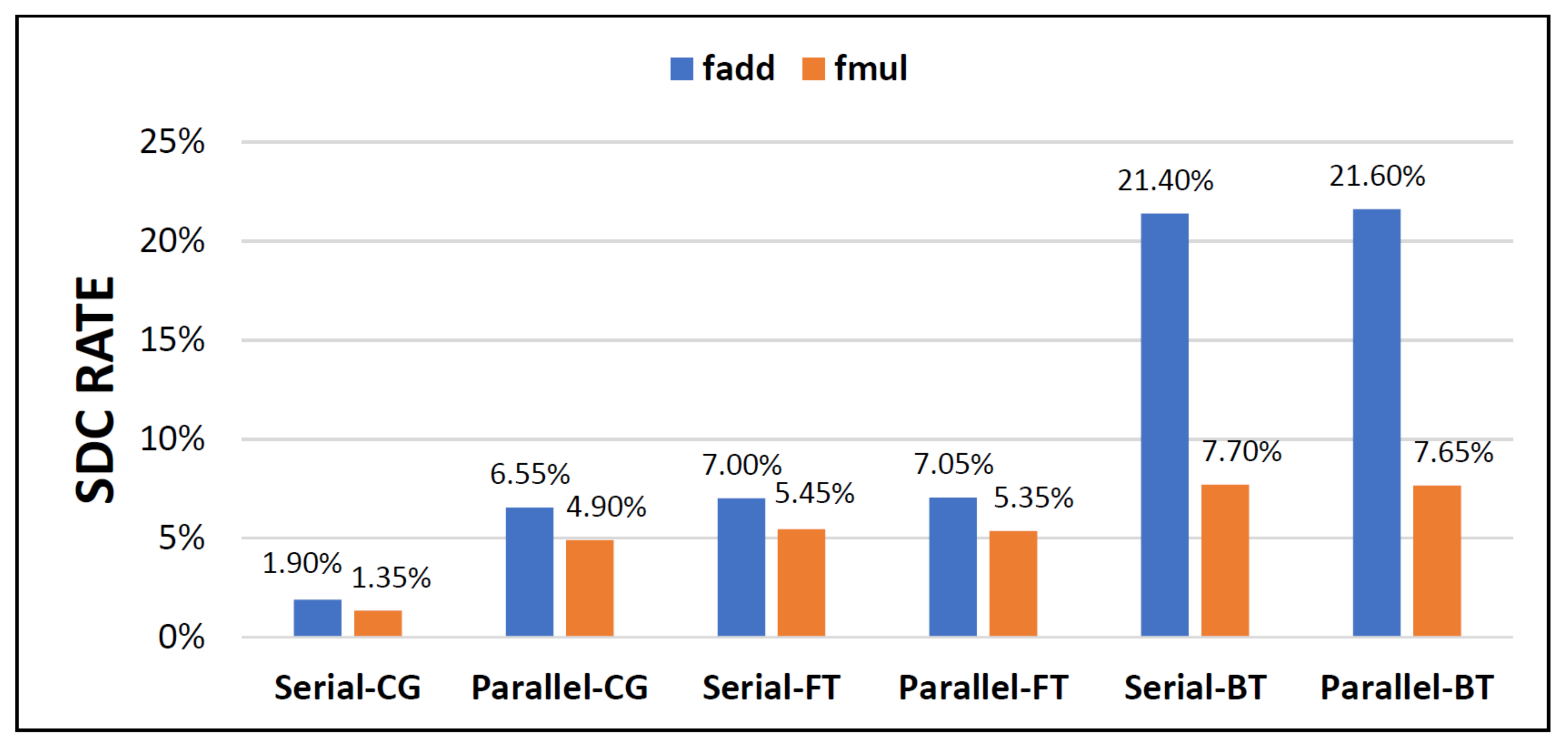}
    \vspace{-13pt}
	    \caption{SDC Rate of NPB CG,FT, BT Benchmarks.} 
       \vspace{-20pt}

  \label{fig:sensitity_study_on_num_tests}
   \vspace{-3pt}

\end{figure}

Figure~\ref{fig:sensitity_study_on_num_tests} shows the fault injection results (i.e., SDC rate of fault injection tests). 
We collect 10,000 fault injection test results for each benchmark and calculate the SDC rate every 1000 fault injection tests. 
The fault injection results become stable after first 6,000 tests.

\begin{figure*}[t]
  \centering
  
   \subcaptionbox{FT (Serial + Benign)}[.23\linewidth][c]{%
    \includegraphics[width=.23\linewidth]{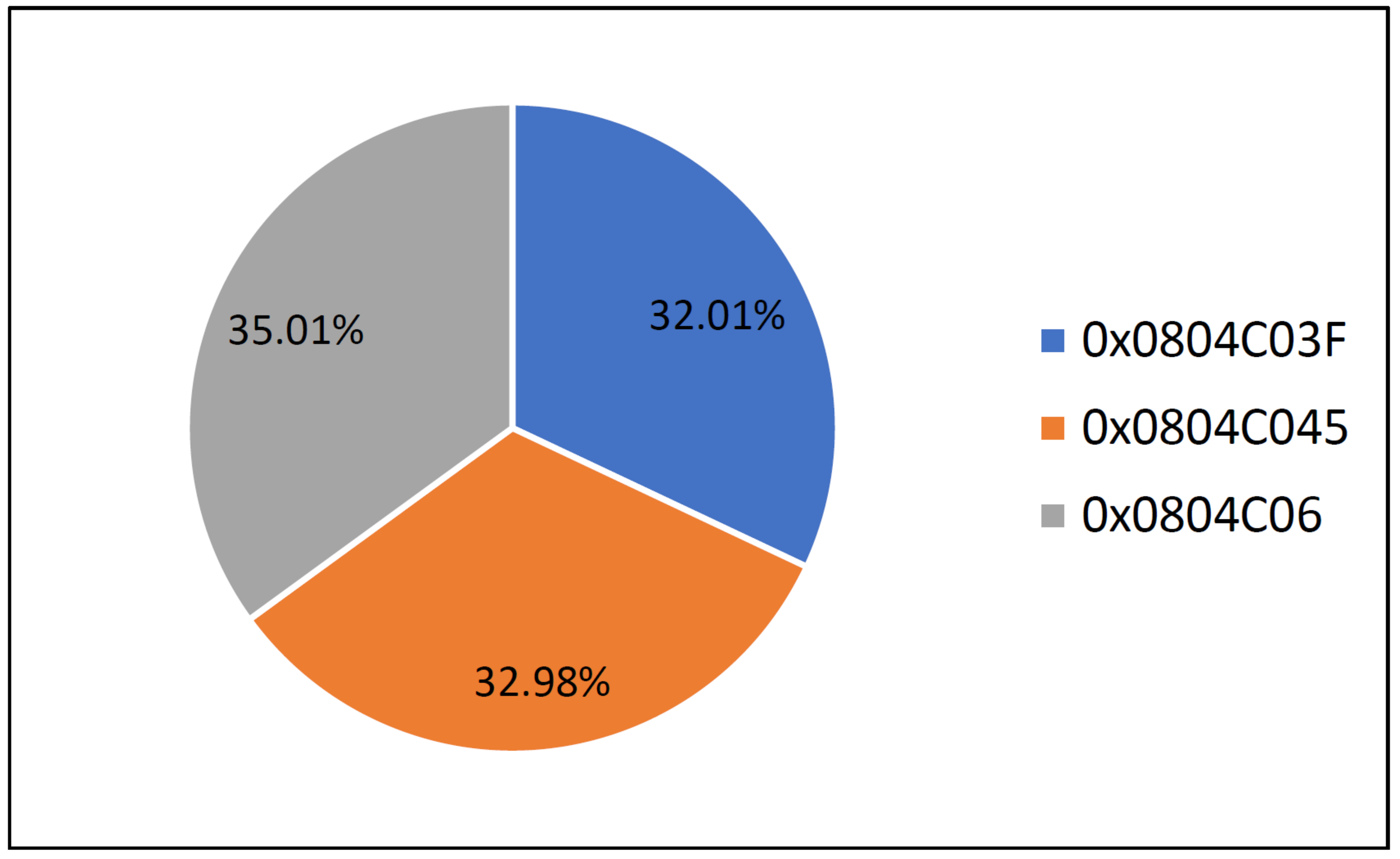}}\quad
  \subcaptionbox{FT (Serial + SDC)}[.23\linewidth][c]{%
    \includegraphics[width=.23\linewidth]{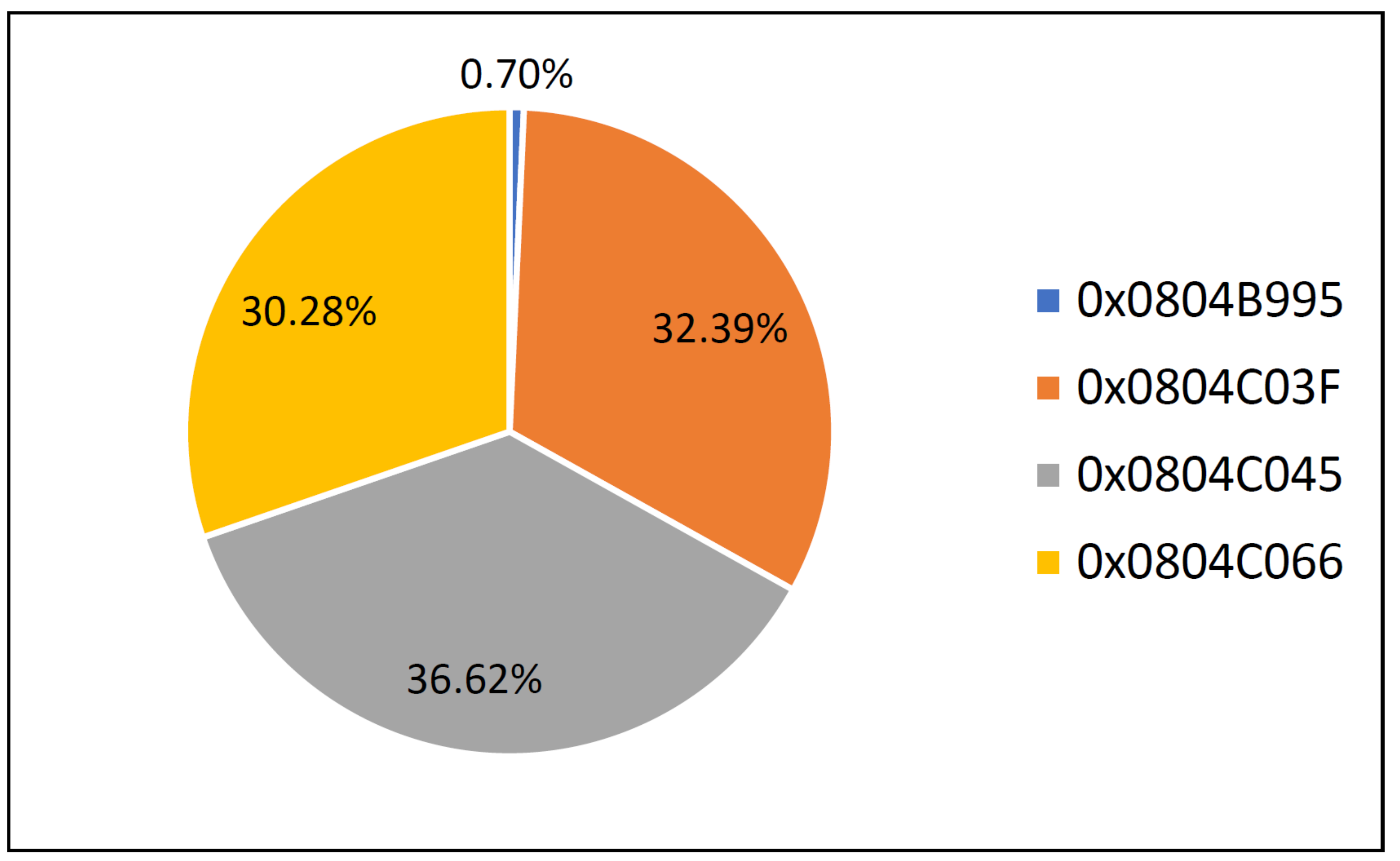}}\quad
  \subcaptionbox{FT (Parallel + Benign)}[.23\linewidth][c]{%
    \includegraphics[width=.23\linewidth]{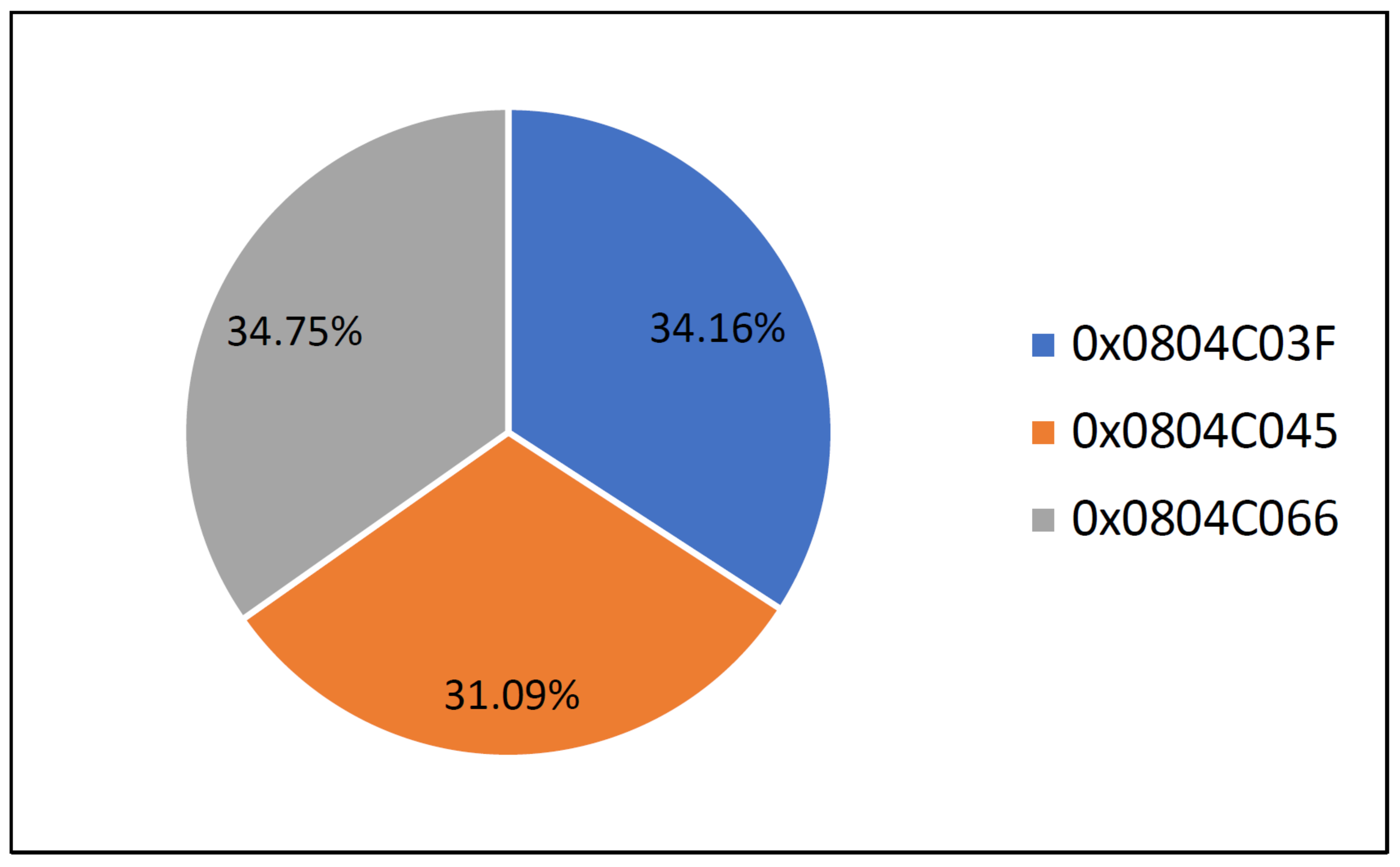}}\quad
  \subcaptionbox{FT (Parallel + SDC)}[.23\linewidth][c]{%
    \includegraphics[width=.23\linewidth]{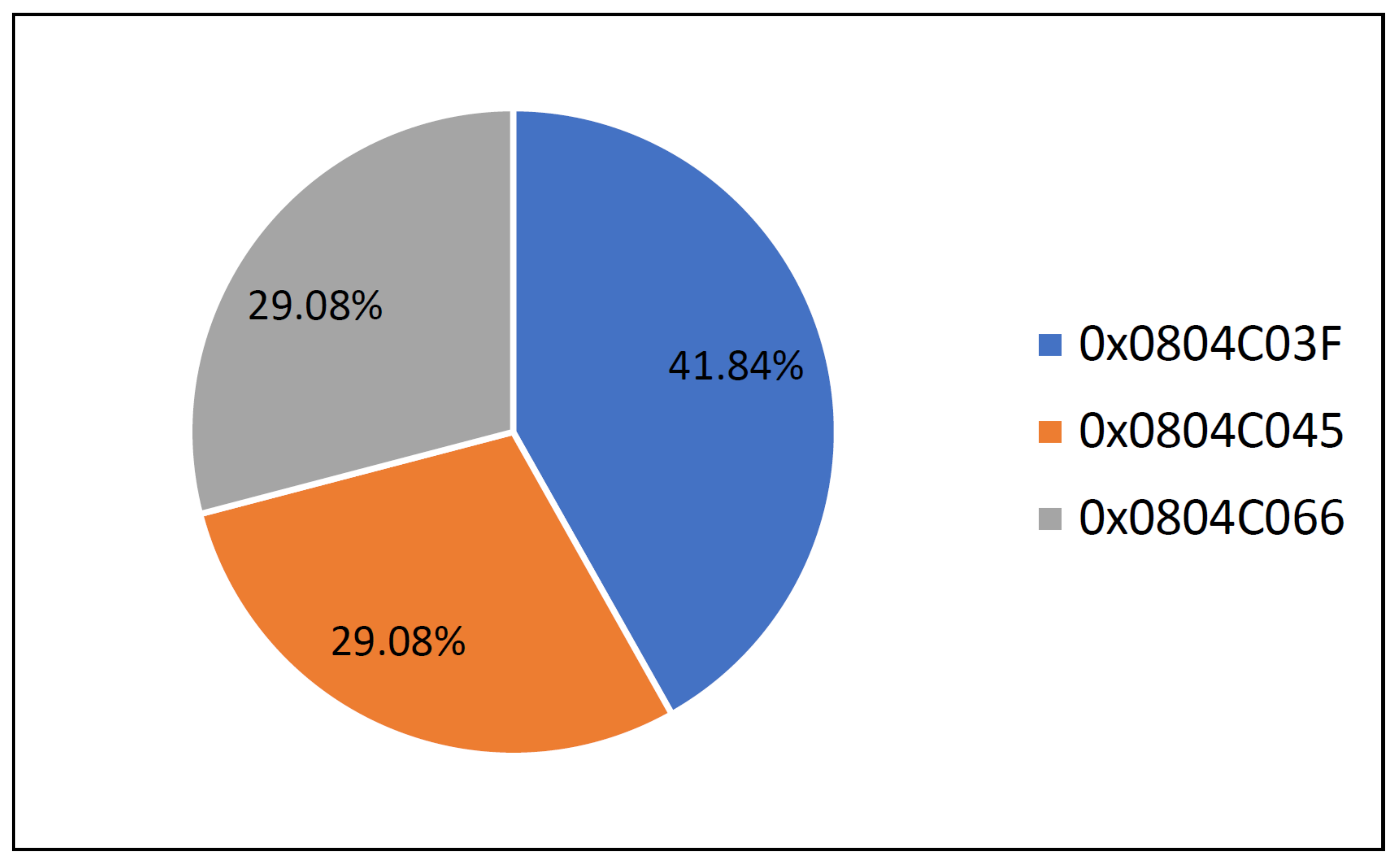}} 
  
     \vspace{-3pt}
   \subcaptionbox{BT (Serial + Benign)}[.23\linewidth][c]{%
    \includegraphics[width=.23\linewidth]{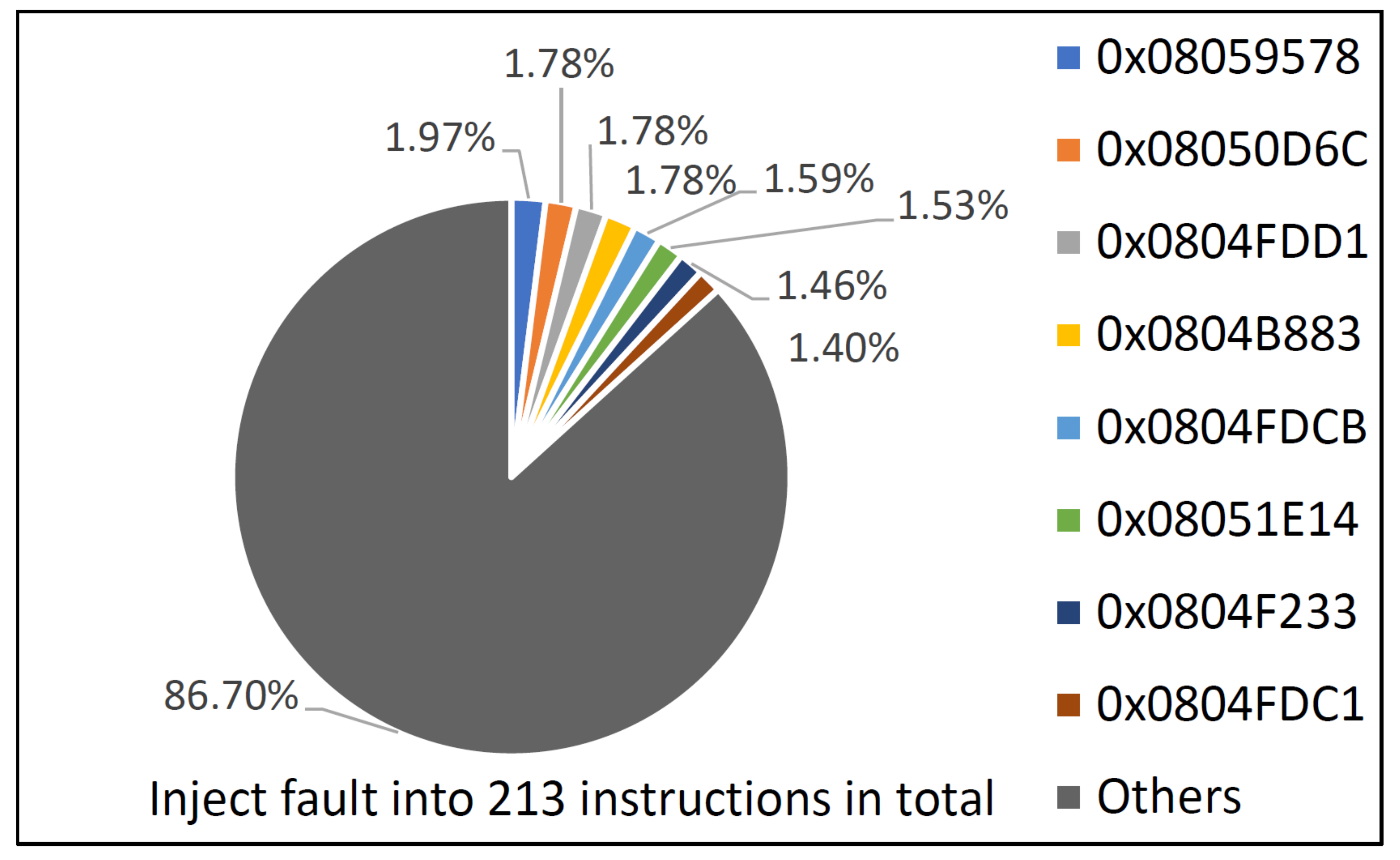}}\quad
  \subcaptionbox{BT (Serial + SDC)}[.23\linewidth][c]{%
    \includegraphics[width=.23\linewidth]{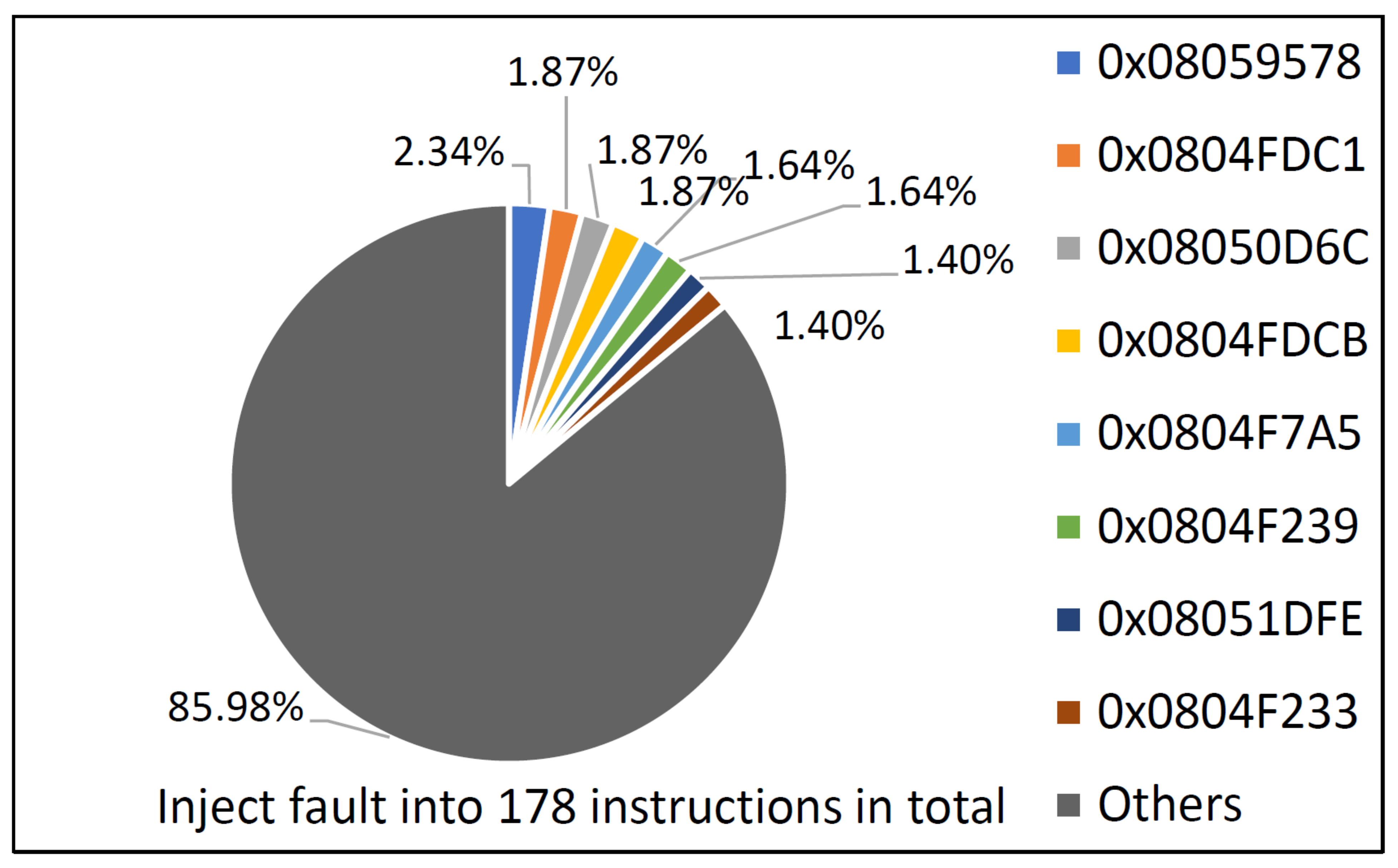}}\quad
  \subcaptionbox{BT (Parallel + Benign)}[.23\linewidth][c]{%
    \includegraphics[width=.23\linewidth]{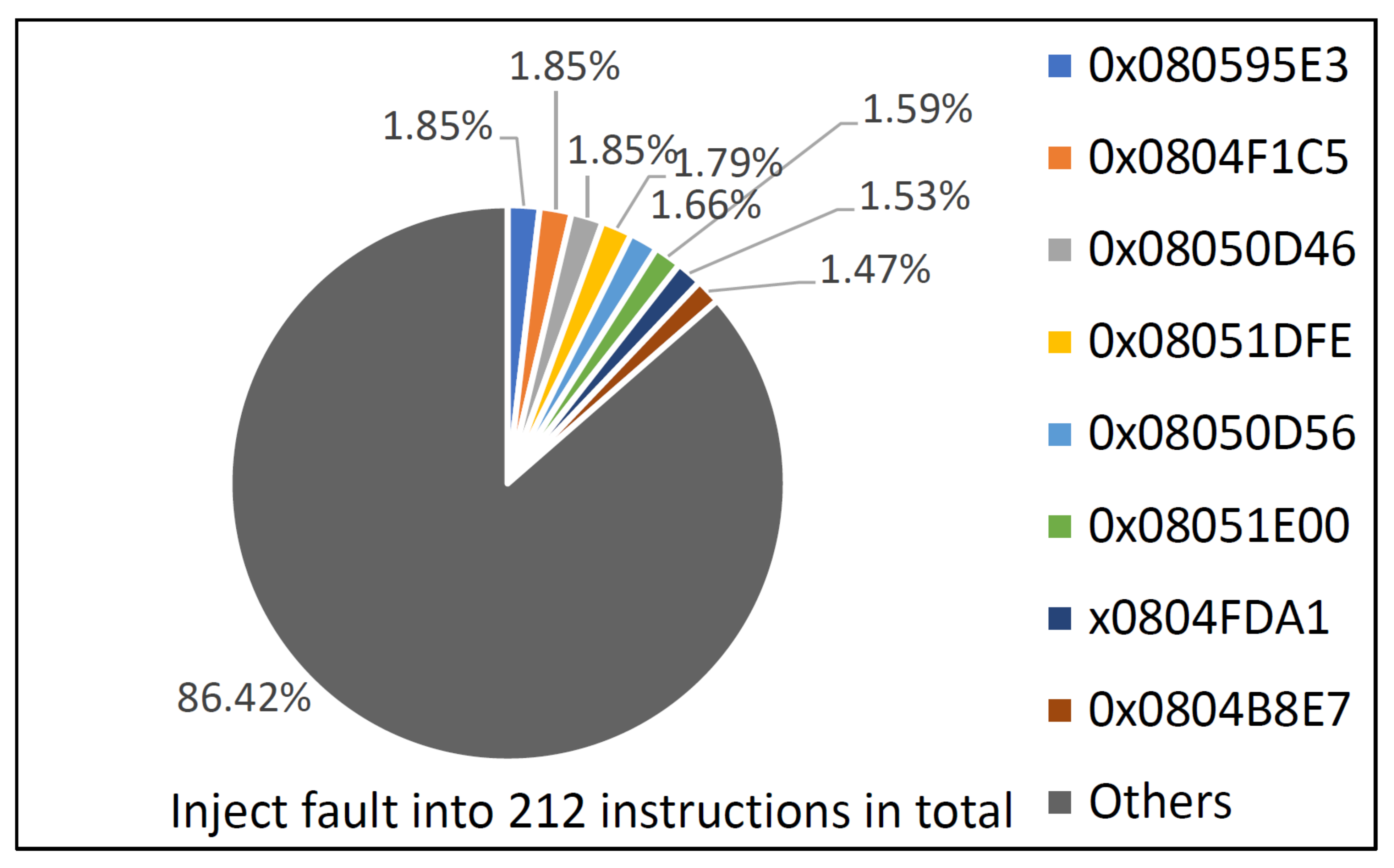}} \quad
  \subcaptionbox{BT (Parallel + SDC)}[.23\linewidth][c]{%
   \includegraphics[width=.23\linewidth]{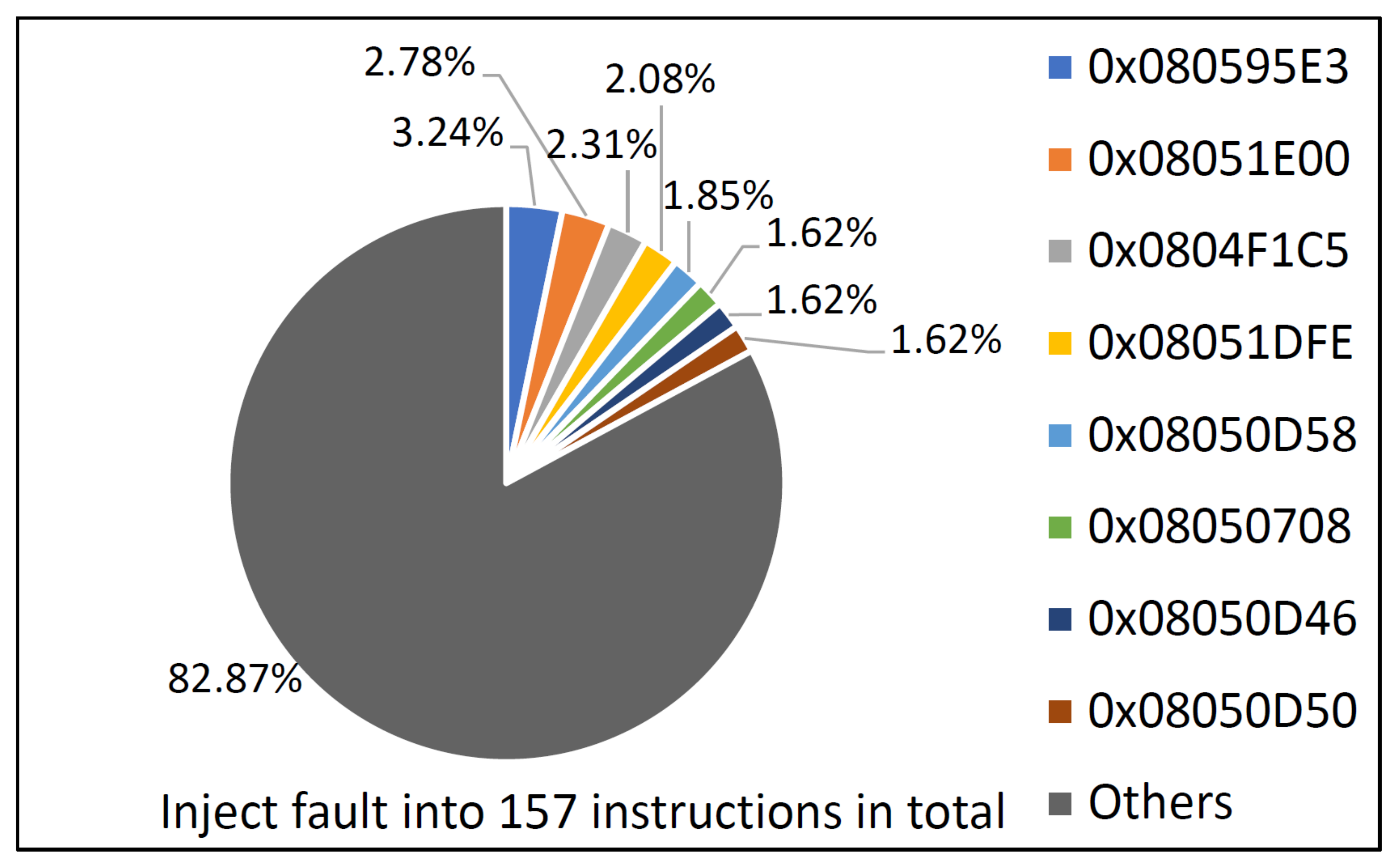}} 

     \vspace{-3pt}
  
  \subcaptionbox{CG (Serial + Benign)}[.23\linewidth][c]{%
   \includegraphics[width=.23\linewidth]{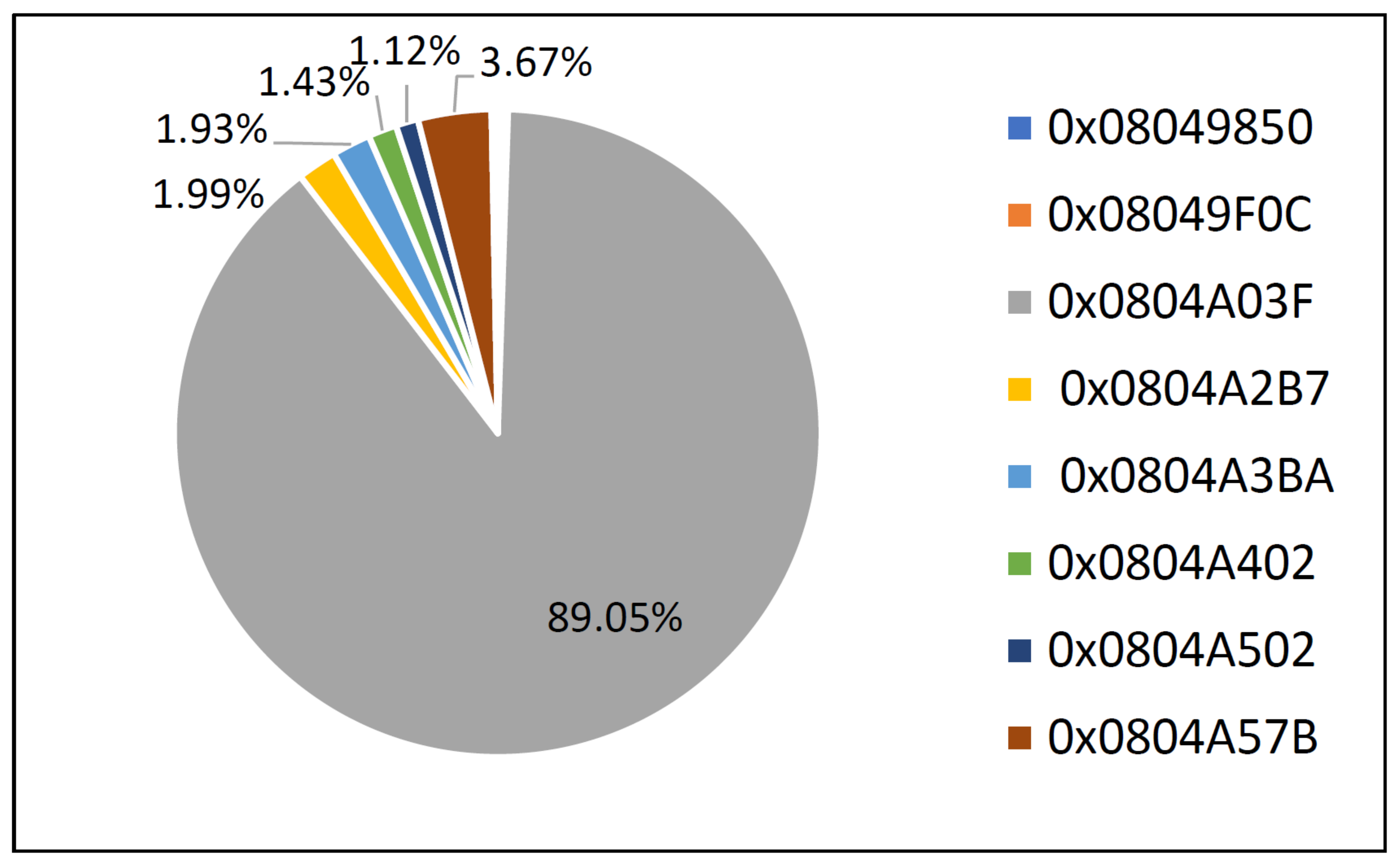}}\quad
  \subcaptionbox{CG (Serial + SDC)}[.23\linewidth][c]{%
  \includegraphics[width=.23\linewidth]{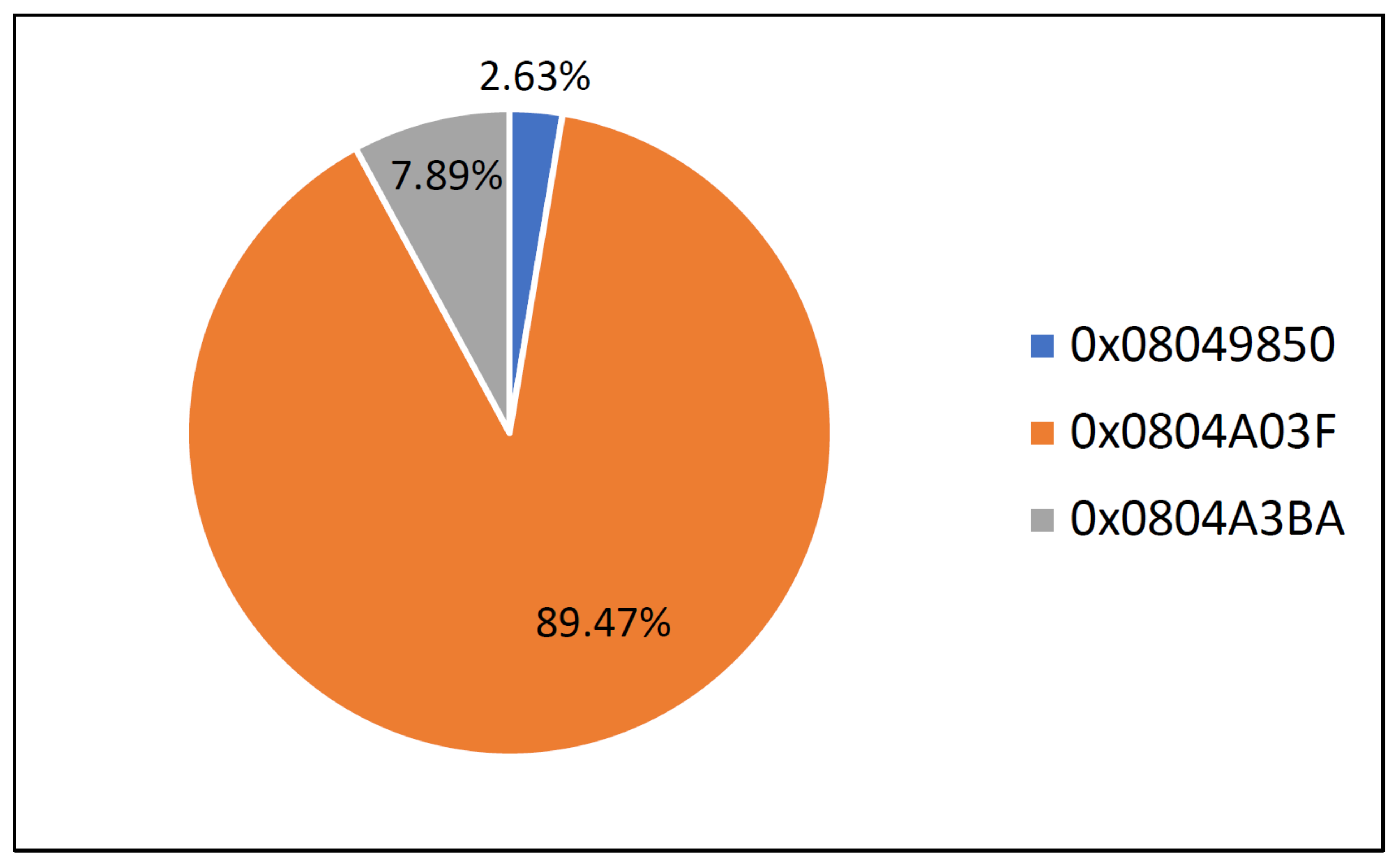}}\quad
  \subcaptionbox{CG (Parallel + Benign)}[.23\linewidth][c]{%
  \includegraphics[width=.23\linewidth]{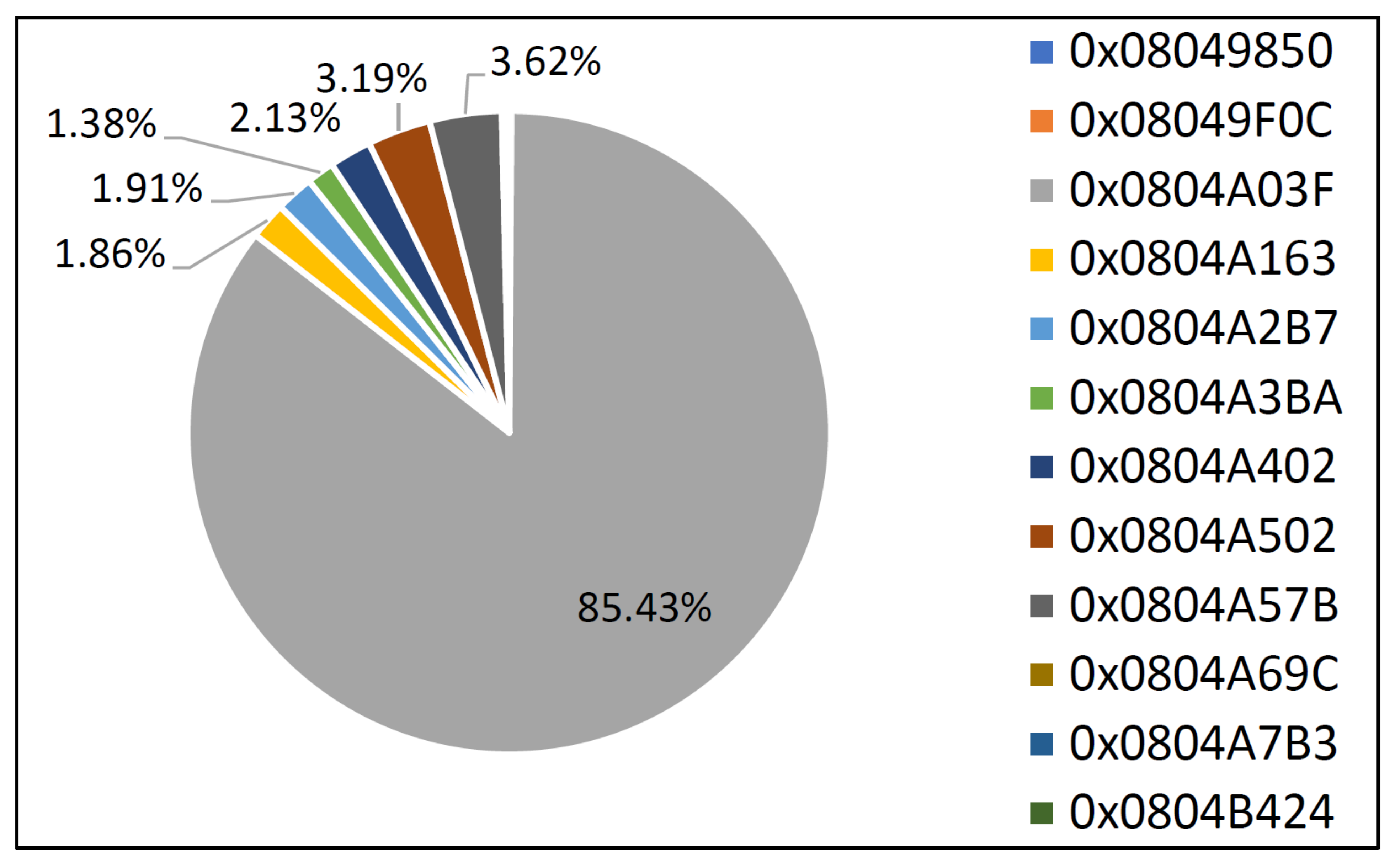}}\quad
  \subcaptionbox{CG (Parallel + SDC)}[.23\linewidth][c]{%
  \includegraphics[width=.23\linewidth]{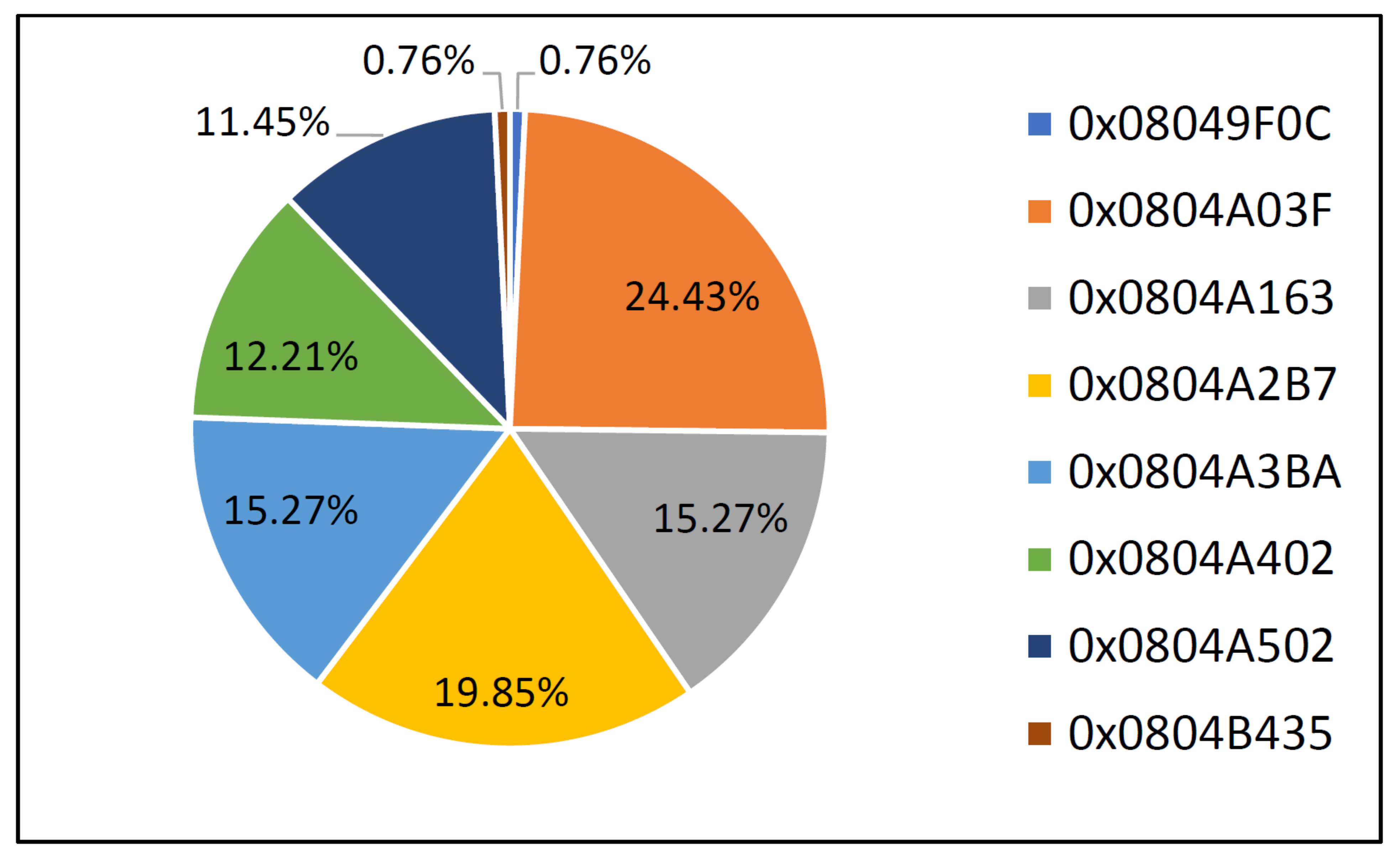}}
  
  \vspace{-13pt}
  \caption{The distribution of faulty floating point \textit{add} instructions in fault injection tests.}
     \vspace{-12pt}
    \label{fig:instr_dist}
      \vspace{-5pt}
\end{figure*}

Figure~\ref{fig:instr_dist} shows the faulty instruction distribution for the fault injection tests on floating point \textit{add} instructions. 
In particular, we find that there are no crashes happened in tests of three benchmarks; thus we only show how frequent each instruction is selected when the fault injection results are benign and SDC.

Figure~\ref{fig:instr_dist} (a)-(d)shows that for FT, the randomly selected faulty instructions in the fault injection tests for the serial and parallel executions are the same, which explains why the fault injection results for the two executions are almost the same.
The fact that the faulty instructions are the same 
mainly because of the code similarity between the serial and parallel codes. 

For BT(see figure~\ref{fig:instr_dist}(e)-(h)), we find that faulty instructions are widely spread across the parallel and serial executions. There is almost no instruction similarity in those faulty instructions between the serial and parallel executions.
It is because BT has complicated computation. There is no dominant computation phase where the faulty instructions can repeatedly happen. 

For CG(see figure~\ref{fig:instr_dist}(i)-(l)), we find that faulty instructions are limited to a few instructions, which is very different from the cases of BT. Also, the fault injection results for the serial and parallel executions are quite different. To understand the reason for such difference, we map the faulty instructions into the source code of CG and have the following observations. 




\vspace{-3pt}

\textbf{Observation 1}: 
The instruction at 0x0804A03F is the most frequently selected instruction for fault injection. Such instruction appears in all cases (serial+benign), (serial+SDC),(parallel+benign) and (parallel+SDC). 
This instruction is used so often in the benchmark, such that most of faults are injected into it. Also, the corruption of this instruction seems to easily cause SDC.
\vspace{-3pt}

\begin{figure}[!t]
    \centering
    \includegraphics[width=0.45\textwidth, height=0.13\textheight]{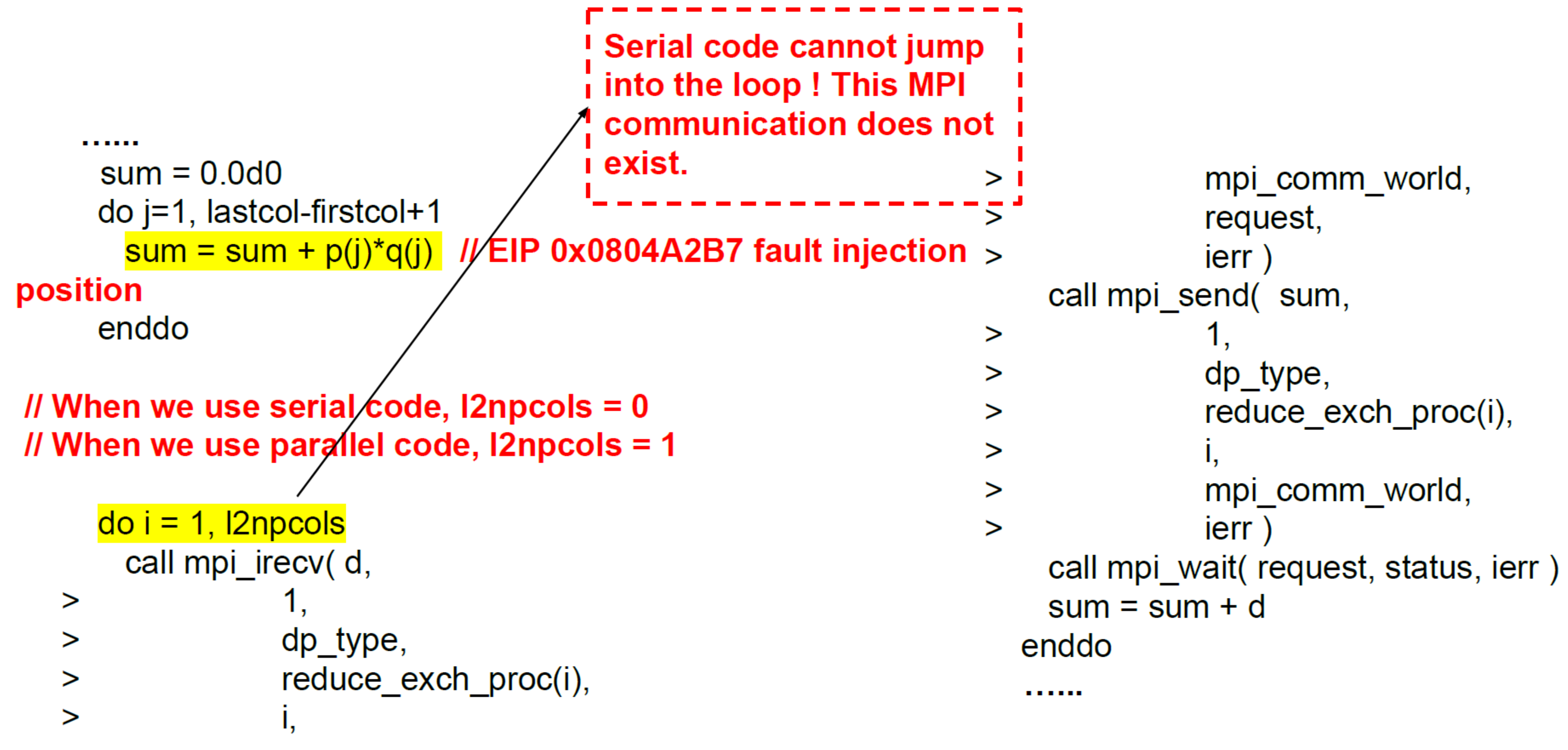}
    \vspace{-15pt}
	    \caption{Source code analysis for the observation 2} 
\vspace{-15pt}
		\label{fig:ob2}
\vspace{-2pt}
\end{figure}

\textbf{Observation 2}:  
Some instructions only appear in (serial+benign), (parallel+benign) and (parallel+SDC), but do not appear in (serial+SDC). Those instructions include those at 0x0804A2B7, 0x0804A3BA, 0X0804A402 and 0x0804A502. 
Those instructions cause fault injection result difference between the serial and parallel executions. 
Figure~\ref{fig:ob2} shows the related code segment for 0x0804A2B7. 
In particular, the serial and parallel executions have a different value for the variable \textit{l2npcols}, which leads to different code structure (particularly the MPI synchronization) for serial and parallel executions. 
Such difference in the code structure makes the faulty injection at 0x0804A2B7 behave differently in the serial and parallel executions.

\vspace{-3pt}

\begin{figure}[!t]
    \centering
    \includegraphics[width=0.45\textwidth, height=0.13\textheight]{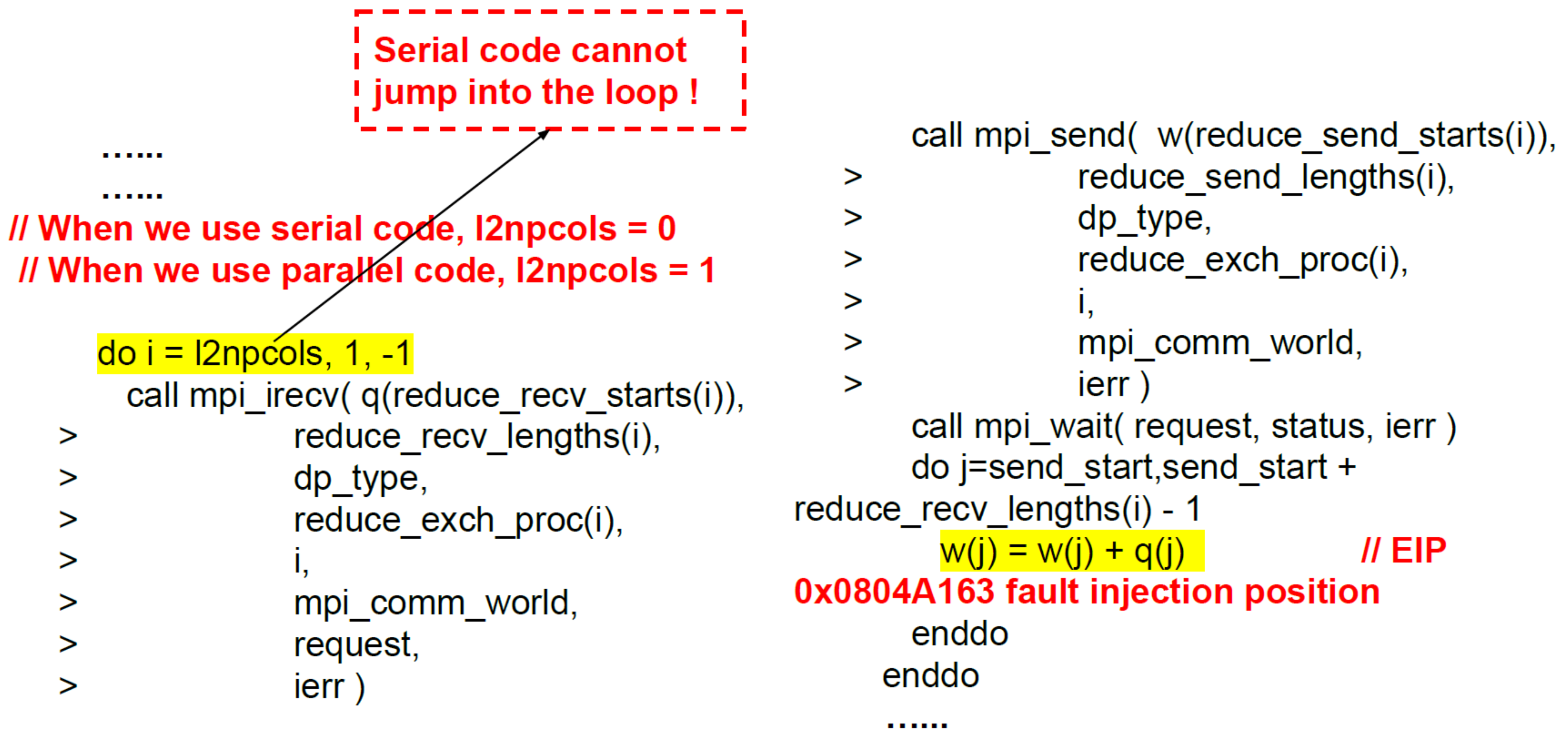}
    \vspace{-15pt}
	    \caption{Source code analysis for the observation 3} 
\vspace{-20pt}
        \label{fig:ob3}
\vspace{-3pt}
\end{figure}

\textbf{Observation 3}: The instruction at 0x0804A163 is only shown in (parallel+bengin) and (parallel+SDC), and such instruction only exists in the parallel execution because of the following reason: the variable \textit{l2npcols} has a different value in the serial and parallel executions. Hence the two executions behave differently (Figure~\ref{fig:ob3}).

%% file: text/conclusions.tex
\vspace{-13pt}
\section{Conclusions and Future Work}
   \vspace{-5pt}
This work is a preliminary study to explain the reason for similar or different application resilience between the serial and parallel executions.
For the future work, we will investigate more benchmarks and establish a model to predict application resilience for the parallel execution based on the fault injection results for the serial execution.
   \vspace{-10pt}

%% file: paper.bbl
\begin{thebibliography}{2}
\providecommand{\natexlab}[1]{#1}
\providecommand{\url}[1]{\texttt{#1}}
\expandafter\ifx\csname urlstyle\endcsname\relax
  \providecommand{\doi}[1]{doi: #1}\else
  \providecommand{\doi}{doi: \begingroup \urlstyle{rm}\Url}\fi

\bibitem[pye()]{pyelftool:github}
{PYELFTOOLS}.
\newblock \url{https://github.com/eliben/pyelftools}.

\bibitem[Guan et~al.(2016)Guan, BeBardeleben, Wu, Eidenbenz, Blanchard, Monroe,
  Baseman, and Tan]{PFSEFI:SIMUTOOLS16}
Q.~Guan, N.~BeBardeleben, P.~Wu, S.~Eidenbenz, S.~Blanchard, L.~Monroe,
  E.~Baseman, and L.~Tan.
\newblock Design, use and evaluation of p-fsefi: A parallel soft error fault
  injection framework for emulating soft errors in parallel applications.
\newblock In \emph{the 9th EAI International Conference on Simulation Tools and
  Techniques}, Prague, Czech Republic, 2016.

\end{thebibliography}
